\documentclass[prb,preprint,showpacs,preprintnumbers,amsmath,amssymb]{revtex4}

\usepackage[dvips]{graphicx}
\usepackage{dcolumn}
\usepackage{bm}
\usepackage{times}
\usepackage{mathptmx}
\usepackage{color}

\begin{document}


\title{Giant third-order magneto-optical rotation in ferromagnetic EuO}

\author{Masakazu Matsubara$^{1,2}$}
 \email{masakazu.matsubara@mat.ethz.ch}
\author{Andreas Schmehl$^{3}$}
\author{Jochen Mannhart$^{4}$}
\author{Darrell G. Schlom$^{5}$}
\author{Manfred Fiebig$^{1,2}$}
 \affiliation{$^{1}$Department of Materials, ETH Zurich, Wolfgang-Pauli-Strasse 10, 8093 Zurich, Switzerland}
 \affiliation{$^{2}$HISKP, Universit\"{a}t Bonn, Nussallee 14-16, 53115 Bonn, Germany}
 \affiliation{$^{3}$Institut f\"{u}r Physik, Universit\"{a}t Augsburg, Augsburg 86135, Germany}
 \affiliation{$^{4}$Max Planck Institute for Solid State Research, Heisenbergstra{\ss}e 1, 70569 Stuttgart, Germany}
 \affiliation{$^{5}$Department of Materials Science and Engineering, Cornell University, Ithaca, New York 14853-1501, USA}

\date{\today}

\begin{abstract}
A magnetization-induced rotation in the third-order nonlinear optical response is observed in out-of-plane-magnetized epitaxial EuO films. We discuss the relation of this nonlinear magneto-optical rotation to the linear Faraday rotation. It is allowed in all materials but, in contrast to the linear Faraday rotation, not affected by the reduction of the thickness of the material. Thus, the third-order magneto-optical rotation is particularly suitable for probing the magnetization of functional magnetic materials such as ultra-thin films and multilayers.

\end{abstract}



\maketitle


\section{Introduction: Nonlinear magneto-optics} \label{intro}

Michael Faraday's discovery of magnetically induced optical activity in 1846\cite{Faraday_1846_PTRS} constituted the first conclusive demonstration of an intimate connection between light and magnetism. This so-called Faraday effect exists in {\it all} media and has long been applied to study the magnetic and electronic properties of materials and image magnetic domain structures.\cite{Zvezdin_1997_MMMM, Sugano_2000_MO} By controlling the polarization of light, a key functionality in modern opto-technology, the Faraday effect plays a crucial role in applications such as optical rotators, isolators, modulators, and circulators.\cite{Zvezdin_1997_MMMM, Sugano_2000_MO}

During the past two decades, nonlinear optical effects, such as sum and difference frequency generation, entered the realm of magneto-optics.\cite{Bennemann_1998_NOM, Fiebig_2005_JOSAB} With nonlinear optics, unique information about the crystallographic, geometric, electronic, and magnetic structure can be acquired. It often addresses states that are inaccessible by linear optics so that a search for the nonlinear analogues of the established linear magneto-optical effects commenced.

Thus far, the vast majority of investigations is focused on second-order magneto-optical effects like magnetically induced second-harmonic generation (SHG).\cite{Bennemann_1998_NOM, Fiebig_2005_JOSAB} Since SHG in the electric-dipole approximation is limited to systems without center of inversion, it is particularly valuable for investigating the inherently noncentrosymmetric surface or interface in centrosymmetric magnets.\cite{Bennemann_1998_NOM} A rotation of the polarization of a reflected SHG wave with respect to the polarization of the incident fundamental light wave, the so-called ``nonlinear magneto-optical Kerr effect'', was reported.\cite{Bennemann_1998_NOM, Pustogowa_1994_PRB, Koopmans_1995_PRL, Rasing_1997_JMMM} The nonlinear Kerr rotation can be orders of magnitude larger than the linear Kerr rotation since, in contrast to the linear case, the magnetization-induced nonlinear contributions to the susceptibility tensor can be of the same order of magnitude as the magnetization-independent ones.

\begin{figure}
\includegraphics[width=\columnwidth,keepaspectratio,clip]{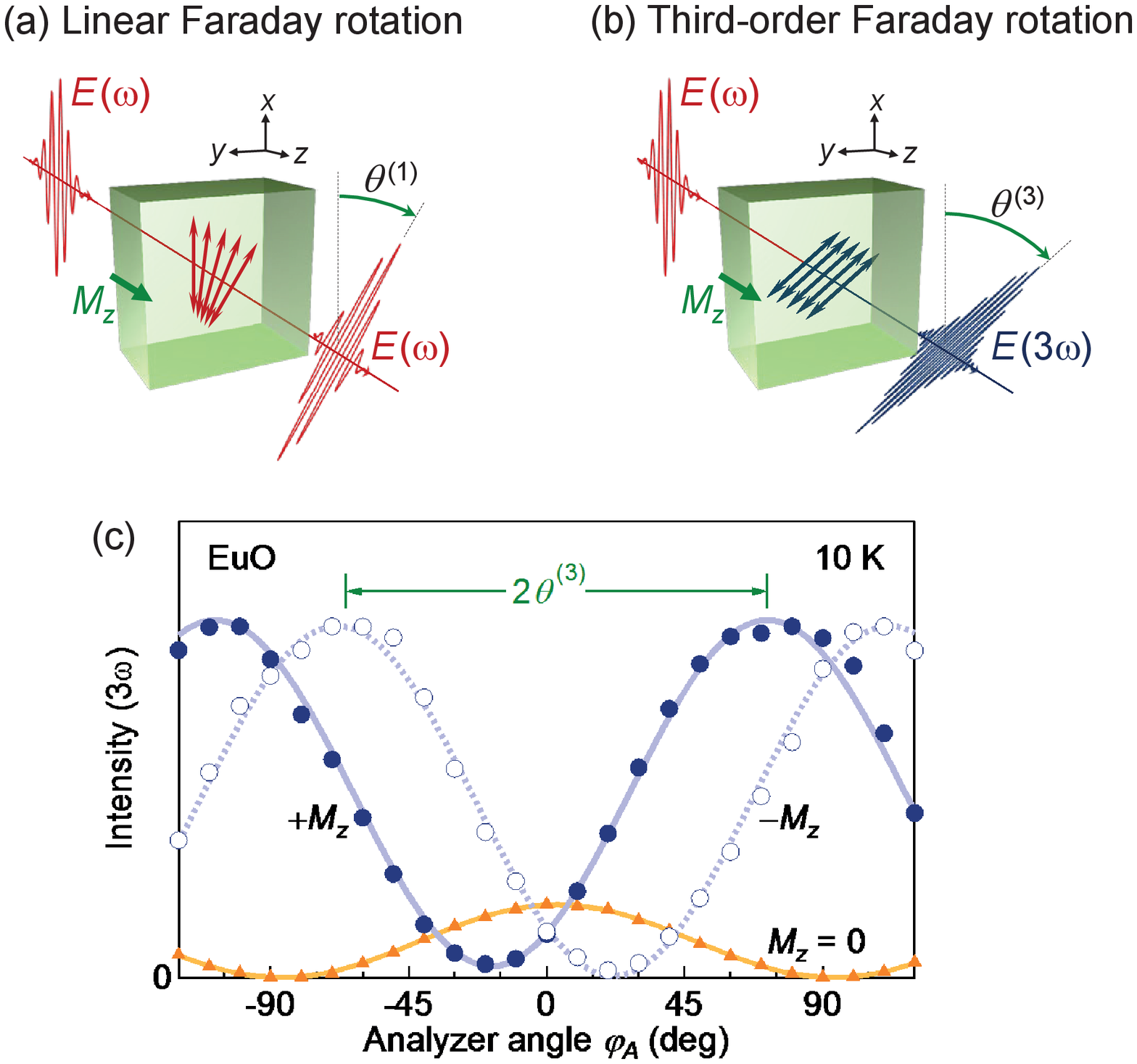}
\caption{(Color online) Schematic of (a) the linear Faraday rotation (LFR) and (b) the third-order Faraday rotation (TFR). The plane of polarization of the outgoing wave at $\omega$ (LFR) or $3\omega$ (TFR) is rotated with respect to the plane of polarization of the ingoing light wave at $\omega$. The rotation is caused by the spontaneous or field-induced magnetization $M_{z}$ parallel to the direction of light propagation $z$. (c) TFR in a EuO(001) film for different $M_{z}$. The measurement shows the intensity of the frequency-tripled light as function of the angular position $\varphi_{A}$ of a linear polarization filter. The nonlinear rotation angle is derived from the value of $\varphi_{A}$ at the maximum of the intensity of the frequency-tripled light. The data for $\pm M_{z}$ and $M_{z}=0$ were obtained in fields of $\pm 3$~T and 0~T applied along the $z$ axis of the EuO(001) film. The lines show sinusoidal fits. The TFR is investigated for $x$-polarized incident light at 10~K and $3\hbar\omega=2.0$~eV.} \label{fig1}
\end{figure}

Yet, the selectivity of SHG can be its major deficiency, because in the majority of magnetically ordered compounds SHG is restricted or even forbidden by symmetry and hence inappropriate for probing their magneto-optical performance. Instead, a nonlinear magneto-optical effect {\it unrestricted} by symmetry is called for. Here an extension of the linear Faraday rotation (LFR) into the regime of harmonic generation is one possibility. Taking third-harmonic generation (THG) as an example, the principle of such a higher-order effect is shown in Fig.~\ref{fig1}(b) in comparison to the LFR depicted in Fig.~\ref{fig1}(a). Both effects correspond to a rotation of the plane of polarization of the emitted light (frequency $n\omega$ with $n=1, 3$) with respect to the polarization of the incident light (frequency $\omega$) by an angle $\theta^{(n)}$. The rotation is generated by the spontaneous or magnetic-field-induced magnetization of the sample along the direction of light propagation. Because of this similarity it is intuitive to consider the process shown in Fig.~\ref{fig1}(b) as higher-order Faraday rotation. However, such a denomination first needs to be justified by placing the nonlinear magneto-optical rotation and the LFR on a common basis, macroscopically as well as microscopically.

\section{Magneto-optical rotation of polarization}

For identifying a common macroscopic basis for the two processes depicted in Figs.~\ref{fig1}(a) and \ref{fig1}(b), we will first review the equations leading to the LFR. Then the formalism will be expanded to the regime of harmonic generation. We will see that the third-order rotation, i.e., the rotation of polarization of the frequency-tripled light wave with respect to the polarization of the incident fundamental light wave (see Fig.~\ref{fig1}(b)), has many properties in common with the LFR and can therefore be interpreted as its nonlinear complement. For simplicity we restrict the discussion to isotropic and uniaxial media in the absence of linear gyrotropy and absorption so that the magneto-optical rotation does not interfere with other dichroic and birefringence effects. The direction of the magnetization $M$ is chosen along the high-symmetry $z$ axis.

\subsection{Linear magneto-optical rotation: The Faraday rotation} \label{Faraday}

The LFR, expressed by the magneto-optical rotation of the linear polarization of light at the frequency $\omega$ propagating through a material in the direction parallel to that of $M$, is derived by inserting the linear dielectric tensor
\begin{equation}
 \hat{\epsilon} = \left(\begin{array}{ccc}
\epsilon_{\|}      & -\epsilon_{\bot}(M) & 0                    \\
\epsilon_{\bot}(M) & \epsilon_{\|}       & 0                    \\
0                  & 0                   & \epsilon_{\|}^{\prime}
\end{array} \right)
\end{equation}
with $\epsilon_{\|}$ and $\epsilon_{\bot}$ as purely real and imaginary components, respectively, into the wave equation
\begin{equation}
 \left(\frac{\hat{\epsilon}}{c^2}\frac{\partial^2}{\partial t^2}-\bigtriangleup\right)\vec{E}=0
\end{equation}
and solving it for a linearly polarized electromagnetic wave $\vec{E}=\vec{E_{0}}\exp\{-i\omega(t-\frac{n}{c}z)\}$. $\epsilon_{\|}$ and $\epsilon_{\bot}(M)\propto M$ denote the elements of the linear dielectric function describing the propagation of light polarized parallel and perpendicular, respectively, to the polarization of the incident light. In general, the off-diagonal component $\epsilon_{\bot}$ is much smaller than the diagonal component $\epsilon_{\|}$. We obtain two eigenmodes, for the electromagnetic wave transmitting through the material, represented by $n_{\pm}^2=\epsilon_{\|}\mp{\rm i}\epsilon_{\bot}$ with $n_{+}$ and $n_{-}$ as refractive index of light with right- and left-handed circular polarization, respectively. For the geometry in Fig.~\ref{fig1}(a) the plane of polarization of the incident linearly polarized light is rotated by the angle
\begin{equation}
\theta_{F} = -\frac{\omega}{c}\Delta n\ell,
\label{eq:rot-1}
\end{equation}
where $\Delta n=(n_{+}-n_{-})/2$ and $\ell$ is the length of the light path in the material along the direction of $M$. Thus, the LFR arises due to the magnetization-induced circular birefringence and $\theta_{F}$ is proportional to the thickness of the material.

\subsection{Non-linear magneto-optical rotation} \label{NOMOR}

In analogy to the definition of the nonlinear magneto-optical Kerr effect we can now introduce the $n$-th-order magneto-optical rotation as rotation of the harmonic wave at $n\omega$ with respect to the polarization of the incident fundamental light wave at $\omega$. In the simplest cases this is expressed by
\begin{equation}
 \tan\theta^{(n)} = \frac{{\rm i}\epsilon_{\bot}^{(n)}(M)}{\epsilon_{\|}^{(n)}}\,,\quad n\geq 2.
\label{eq:rot-n}
\end{equation}
with $\epsilon_{\bot}^{(n)}\propto M$ and $\epsilon_{\|}^{(n)}$ as elements of the $n$-th order dielectric function describing the propagation of light polarized perpendicular and parallel, respectively, to the polarization of the incident light.

By inserting into Eq.~(\ref{eq:rot-1}) the definitions of $\Delta n$ and $n_{\pm}$ as given above, Equations~(\ref{eq:rot-1}) and (\ref{eq:rot-n}) can be combined into the general expression
\begin{equation}
 f_{n}(\theta^{(n)}) = {\rm Re}\left( a^{(n)}\frac{{\rm i}\epsilon_{\bot}^{(n)}(M)}{\epsilon_{\|}^{(n)}}\right)\,,\quad n\in \mathbb{N}.
\label{eq:rot}
\end{equation}
with $f_{n}$ and $a^{(n)}$ as a function and a proportionality factor, respectively. We have $f_{1}(u)=u$, $a^{(1)}=(\omega n_{0}/2c)\cdot\ell$ with $n_{0}=(n_{+}+n_{-})/2$ and $f_{n\geq 2}(u)=\tan u$, $a^{(n\geq 2)}=1$. Only for $n=1$ the frequency of the ingoing and the outgoing light is the same which explains the difference in the expressions for $n=1$ and $n\geq 2$. In any case, Eq.~(\ref{eq:rot}) emphasizes that the magneto-optical rotation of any order is determined by the ratio between the off-diagonal and diagonal components of the dielectric tensor of that order. Note that although we neglected absorption thus far, the components of the dielectric tensor can in general be complex. The rotation of the plane of polarization may be therefore accompanied by elliptical contributions.\cite{Zvezdin_1997_MMMM} In Eq.~(\ref{eq:rot}) this is already taken into account by distinguishing between real and imaginary parts.

We now have to identify the nonlinear complement to the LFR by investigating the different orders of $n$. The case $n=1$ leads to the LFR discussed above and shown in Fig.~\ref{fig1}(a). As discussed, the case $n=2$ (as well as $n=4, 6, 8, \dots$) is restricted or even forbidden by symmetry and therefore inappropriate for probing the magneto-optical performance in general. The case $n=3$ is the leading-order nonlinear magneto-optical rotation process that is, like the LFR, allowed in materials of \textit{any} symmetry. The ``nonlinear magneto-optical Kerr effect'' designates the magnetization-induced rotation of polarization of a reflected nonlinear (frequency-doubled) light wave; in exactly the same way we might now use the term ``nonlinear Faraday effect'' for the magnetization-induced rotation of polarization of a transmitted nonlinear (frequency-tripled) light wave with respect to the incident light wave. However, the term ``nonlinear Faraday effect'' is also used for the nonlinear dependence of the LFR on the intensity of the incident light caused by multi-photon absorption.\cite{Frey_1992_JMSJ} For clarity we henceforth employ the term ``third-order Faraday rotation'' (TFR) for the effect discussed in our work. The most obvious difference between the LFR and the TFR is that the former is proportional to the thickness of the material whereas the latter is thickness-independent. The experimental verification of this striking difference will be the topic of section~\ref{Results}.

\begin{table}
\caption{Nonzero elements of the linear and third-order susceptibility tensor relevant to the LFR and the TFR in ferromagnetic EuO. The components are derived by considering $4/\underline{mm}m$ as magnetic point symmetry. Only the experimentally relevant components for a magnetization parallel to the $z$ axis and an irradiation of $x$-polarized fundamental light incident along the $z$ axis are listed.\protect\cite{Birss_1966_SM} Even contributions couple to $M^0$, $M^2$, etc.\ whereas odd contributions couple to $M^1$, $M^3$, etc. In general higher-order terms are so small that only the leading terms coupling to $M^0$ (magnetization-independent) and $M^1$ (linear coupling) need to be considered. A manifestation of higher-order terms will be discussed in section~\ref{Results}.}
\begin{center}
\begin{ruledtabular}
\begin{tabular} {ccc}
 & Even in $M$ & Odd in $M$ \\ \hline
LFR & $\chi_{xx}=\chi_{yy}$ & $\chi_{yx}=-\chi_{xy}$ \\
TFR & $\chi_{xxxx}$ ($=\chi_{yyyy}$) & $\chi_{yxxx}$ ($=-\chi_{xyyy}$) \\
\end{tabular}
\end{ruledtabular}
\end{center}
\label{table1}
\end{table}

For the geometry in Figs.~\ref{fig1}(a) and \ref{fig1}(b), nonzero elements of the linear and third-order susceptibility tensor relevant to the LFR and the TFR are summarized in Table~\ref{table1}. Here the third-order susceptibility is derived from the general expression for THG,
\begin{equation}
P_{i}(3\omega) = \epsilon_{0} \chi^{(3)}_{ijkl} E_{j}(\omega)E_{k}(\omega) E_{l}(\omega).
\label{eq:THG}
\end{equation}
With $\hat{\epsilon}^{(3)}=\hat{\chi}^{(3)}$ in Eq.~(\ref{eq:rot-n}), we obtain
\begin{equation}
\tan\theta^{(3)} = {\rm Re}({\rm i}\chi_{yxxx}(M)/\chi_{xxxx}).
\label{eq:TFR}
\end{equation}
Despite the potential of the TFR as universal magneto-optical probe, only a single study has been reported thus far.\cite{Shimura_2003_APL} In that study, garnet films revealed a rotation of about $4^{\circ}$ and neither the spectral characteristics nor the microscopic origin of the effect were investigated, so that the general aspects of the nature and potential of the TFR remained unclear.

In the following we will show that thin epitaxial films of the ferromagnetic semiconductor EuO display a ``giant'' TFR. The rotation varies between zero in the absence of a magnetic field and about $80^{\circ}$ in a field of 2.5~T. Spectroscopy reveals its microscopic origin. Based on an inherent relation between the TFR and the LFR we point out the general feasibility of the TFR for probing magnetic matter and thin films in particular.

\section{Samples and methods}

\subsection{Ferromagnetic EuO}

EuO is attracting much attention from the point of view of basic science and application.\cite{Schmehl_2007_NatMat} It has a high potential for semiconductor-based spintronics applications\cite{Steeneken_2002_PRL, Schmehl_2007_NatMat, Santos_2008_PRL, Mairoser_2010_PRL} due to its half-metallic behavior with electron doping\cite{Steeneken_2002_PRL, Schmehl_2007_NatMat, Santos_2008_PRL} and its structural and electronic compatibility with Si, GaN, and GaAs.\cite{Schmehl_2007_NatMat, Swartz_2010_APL} At room temperature, stoichiometric EuO is a paramagnetic semiconductor with a band gap of $\sim 1.2$~eV. It orders ferromagnetically at $T_{C}=69$~K. The Eu$^{2+}$ ions have localized $4f^{7}$ electrons with $^{8}S_{7/2}$ as the ground state, yielding a saturation magnetic moment as large as 7~$\mu_{\rm B}$. A multitude of remarkable magneto-optical properties have been revealed in EuO, such as a strong linear and circular birefringence and dichroism,\cite{Freiser_1968_HPA, Ahn_1967_IEEETM, Feinleib_1969_PRL, Wang_1986_HPA} as well as a large well-investigated red-shift of the absorption edge associated to the magnetic ordering.\cite{Freiser_1968_HPA, Matsubara_2010_PRB} With a rotation of $5\cdot 10^{5}$~deg/cm, EuO shows one of the largest LFR.\cite{Ahn_1967_IEEETM} Pronounced magnetization-induced SHG and THG contributions have been observed on the binary Eu compounds and the electronic origin of the SHG and THG spectra has been discussed.\cite{Kaminski_2009_PRL, Matsubara_2010_PRB, Lafrentz_2010_PRB, Matsubara_2011_JAP} Hence, because of outstanding magnetic and optical properties and their strong connection, EuO is an ideal compound for exploring the TFR. Yet, as we will see, the results gained on EuO are instructive for understanding TFR in general.

\subsection{Sample preparation and experimental methods}

Epitaxial EuO(001) films protected by an amorphous silicon (a-Si) cap layer of 10-20~nm were grown by molecular-beam epitaxy on two-side polished YAlO$_{3}$(110) substrates.\cite{Schmehl_2007_NatMat} For most of the measurements, film with a thickness of 100~nm was used. Bulk-like crystallographic, transport, and linear optical properties\cite{Schmehl_2007_NatMat, Matsubara_2010_PRB} confirm the excellent quality of the epitaxial films. Samples were mounted in an optical helium-operated split-coil cryostat in which magnetic fields of up to $\pm 3.5$~T applied along the $z$ axis induced the TFR. The TFR was measured with light incident perpendicular to the EuO surface. The setup for nonlinear transmission spectroscopy is described in detail in Ref.~\onlinecite{Matsubara_2010_PRB}. Light pulses were generated in an optical parametric amplifier pumped by a regenerative Ti:sapphire amplifier system providing a central wavelength of 800~nm (1.55~eV), a pulse width of 120~fs, and a repetition rate of 1~kHz. The TFR was investigated at temperatures of $10-200$~K in the spectral range $3\hbar\omega$ of $1.85-3.50$~eV. The nonlinear spectra of the EuO films were normalized to the reference signal obtained on a wedged $\alpha$-SiO$_{2}$ plate. They were also normalized to the spectral response of the detection system.

\section{Results and discussion} \label{Results}

\subsection{Verifying third-order Faraday rotation}

Figure~\ref{fig1}(c) shows the intensity of the frequency-tripled light as a function of the angular position $\varphi_{A}$ of a polarization filter. Maximum intensity directly reveals the direction of polarization at $3\omega$. Data were taken at 10~K and $3\hbar\omega=2.0$~eV for magnetic fields $\mu_{0}H_{z}$ of 0~T and $\pm 3$~T. At $\mu_{0}H_{z}=0$~T the sample possesses an in-plane magnetization, so that $M_{z}=0$ and $\theta^{(3)}=0^{\circ}$. At $\mu_{0}H_{z}=+3$~T the situation changes entirely. The intensity of the frequency-tripled light is greatly enhanced and its maximum shows a large shift with respect to $\varphi_{A}$. Here the field induces an out-of-plane magnetization $M_{z}\neq 0$ and with it a large rotation of about $70^{\circ}$.

In order to explore the relation of this rotation to the TFR a variety of tests was performed. First, the observed rotation agrees well with the symmetry analysis. In- and out-of-plane magnetized EuO possesses the point symmetry $4m\underline{mm}$ and $4\underline{mm}m$, respectively, and only the latter allows the magnetically induced frequency tripling that can lead to a TFR.\cite{Birss_1966_SM} Second, we note the reversal of the rotation occurring with the reversal of $M_{z}$ in Fig.~\ref{fig1}(c). This is a property required for Faraday rotation of any order. Third, Fig.~\ref{fig2}(a) shows the angular dependence of the frequency-tripled signal as in Fig.~\ref{fig1}(c) for a variety of temperatures in the ferromagnetic and the paramagnetic state. The extracted temperature variation of the rotation is indicated by triangles and entered in Fig.~\ref{fig2}(c) as open squares. We see that in the vicinity of $T_{C}$ the rotation decreases drastically and reflects the decrease of $M_{z}$. Note that the onset temperature of the magnetization in EuO is strongly influenced by external magnetic fields,\cite{Mairoser_2010_PRL, Hayata_2001_JMSJ} which explains the small signal remaining just above $T_{C}$. Contributions by the LFR that may interfere with the third-order rotation are small. At 10~K we find that $\theta^{(1)}(\omega)$ and $\theta^{(1)}(3\omega)$ are $\sim 0^{\circ}$ and $\sim 4^{\circ}$, respectively, in agreement with earlier data.\cite{Ahn_1967_IEEETM}

\subsection{Temperature and magnetic field dependence}

\begin{figure}
\includegraphics[width=\columnwidth,keepaspectratio,clip]{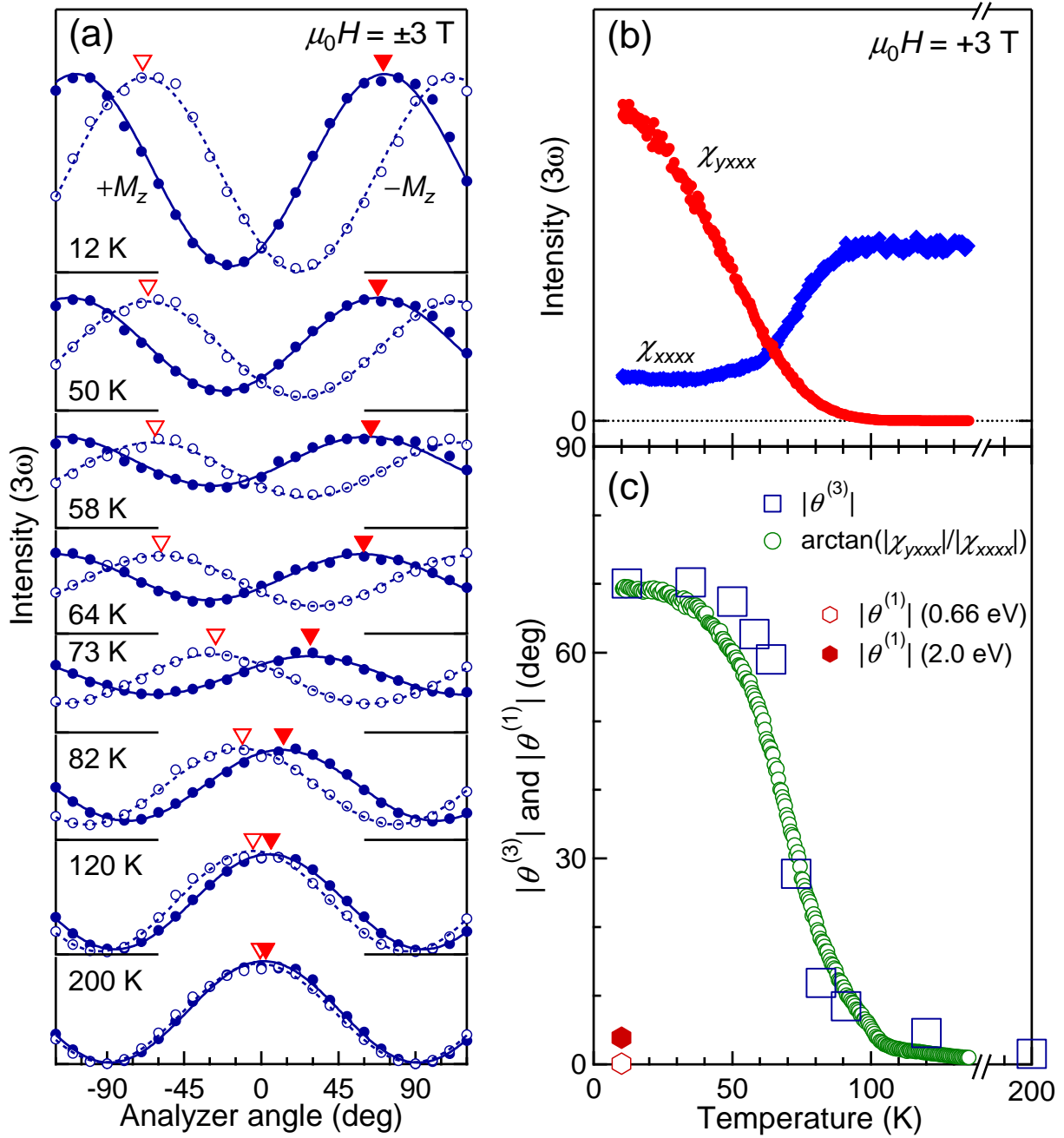}
\caption{(Color online) Temperature dependence of the TFR of a EuO(001) film at
$3\hbar\omega=2.0$~eV. (a) Determination of the nonlinear rotation angle for temperatures between 10~K and 200~K. For $\mu_{0}H_{z}=+3$~T (closed circles) and $\mu_{0}H_{z}=-3$~T (open circles) this angle is indicated by the respective triangles. (b) Temperature dependence of the nonlinear susceptibilities $\chi_{yxxx}$ and $\chi_{xxxx}$ for $\mu_{0}H_{z}=+3$~T. (c) Comparison of the rotation angle derived from (a) (open squares) and (b) (open circles). The hexagons refer to the LFR at 0.66 and 2.0~eV.} \label{fig2}
\end{figure}

\begin{figure}
\includegraphics[width=\columnwidth,keepaspectratio,clip]{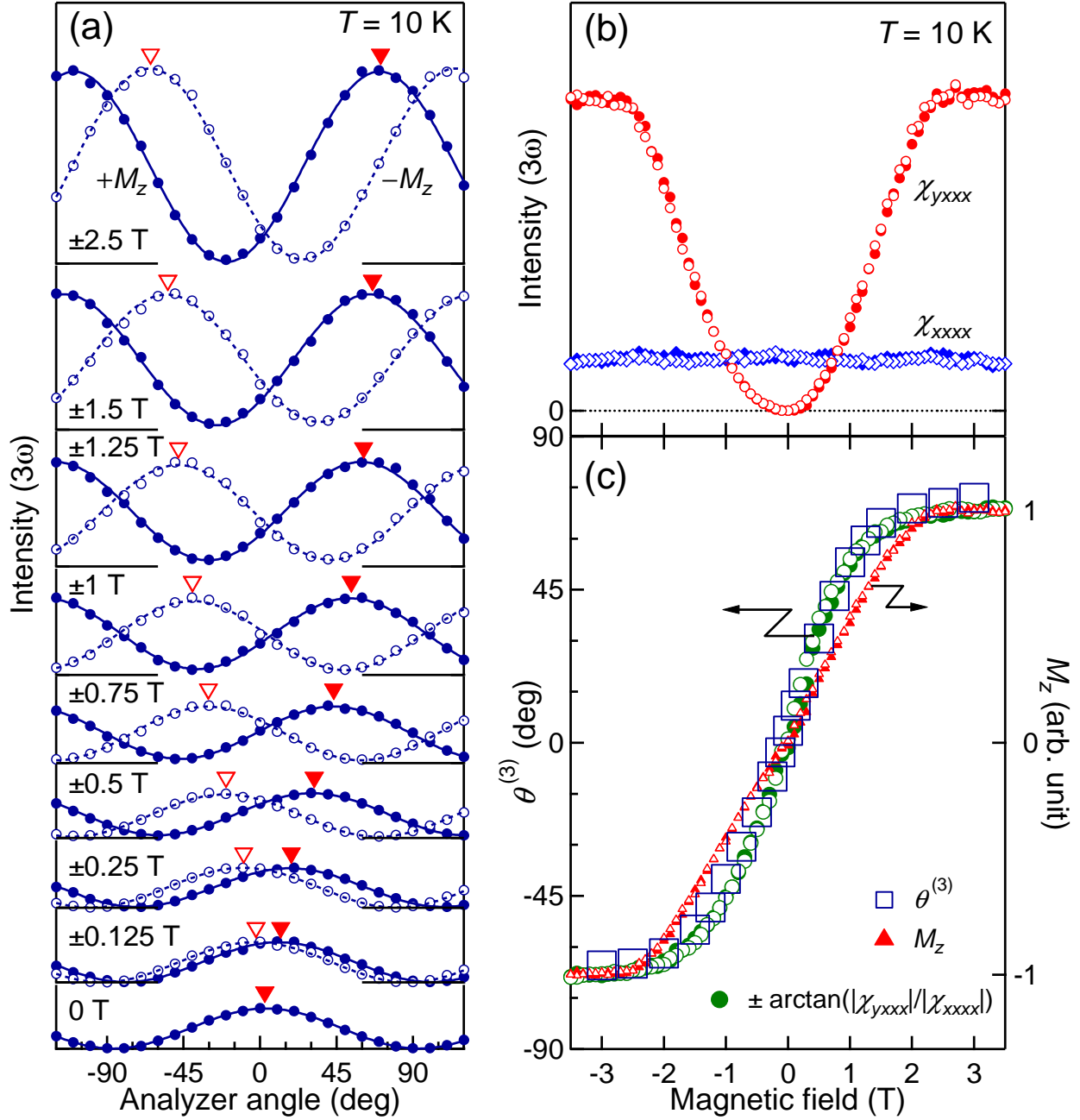}
\caption{(Color online) Magnetic-field dependence of the TFR of a EuO(001) film at $3\hbar\omega=2.0$~eV. (a) Determination of the nonlinear rotation angle as in Fig.~\protect\ref{fig2} for magnetic fields between 0~T and $\pm2.5$~T. (b) Magnetic-field dependence of the nonlinear susceptibilities $\chi_{yxxx}$ and $\chi_{xxxx}$ at 10~K. Data represented by closed (open) symbols were taken with increasing (decreasing) field. (c) Comparison of the rotation angle derived from (a) (open squares) and (b) (closed and open circles). The magnetic-field dependence of the out-of-plane magnetization $M_{z}$ derived from (b) is shown by triangles. It agrees well with published magnetization measurements.\cite{Mairoser_2010_PRL}} \label{fig3}
\end{figure}

With reference to Eq.~(\ref{eq:TFR}), Fig.~\ref{fig2}(b) shows the temperature dependence of the frequency-tripled signal for $\chi_{yxxx}$ and $\chi_{xxxx}$. We find that both susceptibilities change with temperature, in particular around $T_{C}$, albeit in a different way. However, their magnetic-field dependence at a fixed temperature in Fig.~\ref{fig3}(b) reveals that only $\chi_{yxxx}$ responds to the applied field while $\chi_{xxxx}$ does not. With the application of the magnetic field $\chi_{yxxx}$ increases from zero for $M_{z}=0$ to its saturation value at $\geq 2.5$~T. In contrast, $\chi_{xxxx}$ is independent of the applied field and the associated reorientation of the spontaneous magnetization. We therefore see that the variation of $\chi_{xxxx}$ with temperature in Fig.~\ref{fig2}(b) is caused by the large temperature-dependent spectral shift occurring around $T_{C}$.\cite{Freiser_1968_HPA, Matsubara_2010_PRB, Kasuya_1972_CRCCRSS} The coupling to the magnetization is therefore an indirect band-structural effect. Because of the independence of the direction of the magnetization, the band-structural shift may be parametrized by an even-power expansion, yielding in total terms $\propto$ $M_{\rm sat}^2$, $M_{\rm sat}^4$, etc. in $\chi_{xxxx}$ and terms $\propto$ $M_{z} \cdot M_{\rm sat}^2$, $M_{z} \cdot M_{\rm sat}^4$, etc. in $\chi_{yxxx}$) (with $M_{\rm sat}$ as saturation magnetization at a certain temperature). The rotation angles are not directly affected by this band-structural shift because it enters both susceptibilities, $\chi_{xxxx}$ as well as $\chi_{yxxx}$, in the same way (they are probed at the electronic transition, see section~\ref{micro}).

Considering that the frequency-tripled signal $I$ for $\chi_{yxxx}$ is proportional to the square of $M_{z}$ because of $I\propto|\chi|^2$, the dependence of $M_{z}$ on the applied field $H_{z}$ is extracted. The magnetic-field dependence of $\chi_{yxxx}$ in Fig.~\ref{fig3}(c) reproduces the results of earlier measurements of $M_{z}$,\cite{Mairoser_2010_PRL} thus revealing that the coupling of $\chi_{yxxx}$ to $M_{z}$ is indeed linear and in agreement with Table~\ref{table1}. At saturation, $\chi_{yxxx}$ substantially exceeds $\chi_{xxxx}$. This notably contrasts the linear magneto-optical response, where the magnetization-induced susceptibility $\chi_{yx}$ is much smaller than the magnetization-insensitive susceptibility $\chi_{xx}$.

\begin{figure}
\includegraphics[width=\columnwidth,keepaspectratio,clip]{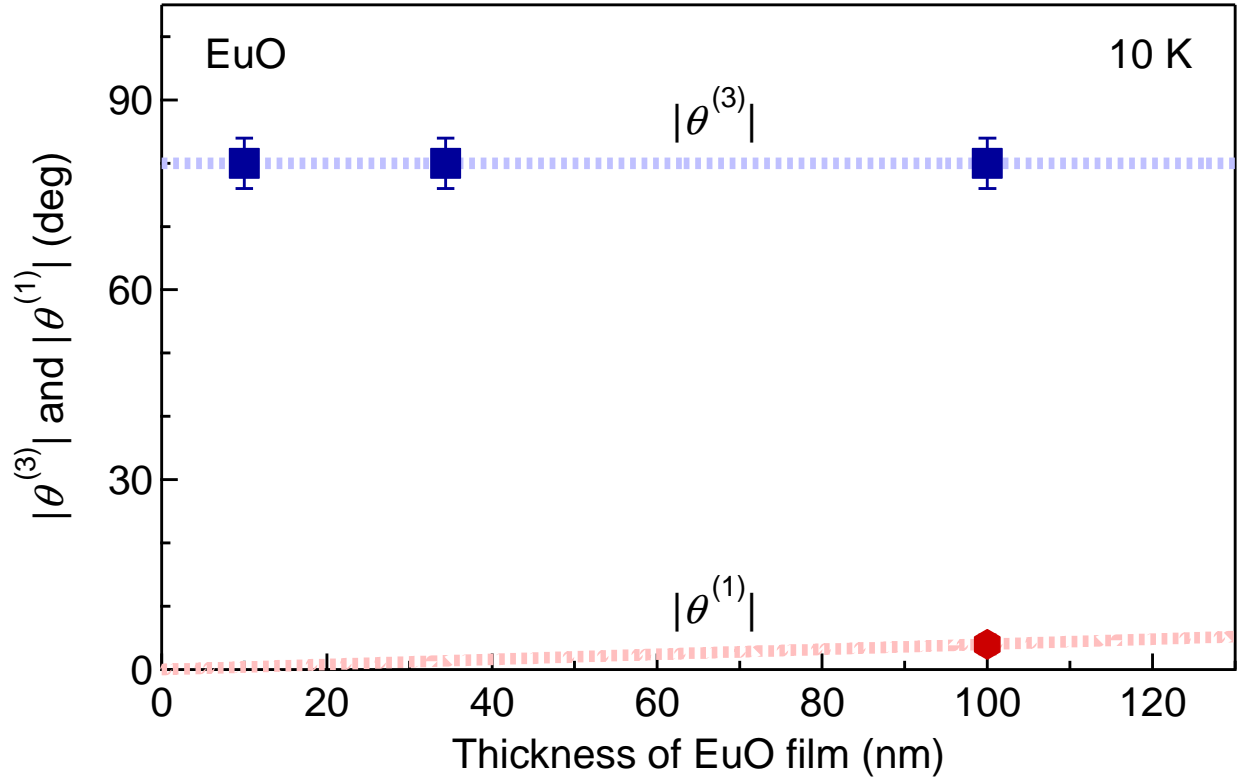}
\caption{(Color online) Thickness dependence of the TFR measured on EuO(001) films at 10~K and $3\hbar\omega = 2.0$~eV. Data were corrected by the measured THG contribution from the a-Si cap layer. The expected linear dependence of the LFR at 2.0~eV is also plotted as a reference.} \label{fig4}
\end{figure}

A noticeable difference distinguishing the TFR from the LFR is the proposed independence of the rotation angle of the thickness of the material. We scrutinized this claim by measuring the TFR for EuO(001) films with a thickness of 100, 34, and 10~nm. Figure~\ref{fig4} shows that within the statistical error the same value $\theta^{(3)}\approx 80^{\circ}$ is observed for all three samples. Thus, TFR can be particularly useful for probing the magnetic properties of very thin films where $\theta^{(1)}$ of the LFR would approach zero. Another distinct difference between the LFR and the TFR is the dependence of the rotation angle on the magnetization. Figure~\ref{fig3}(a) shows the angular dependence of the frequency-tripled signal for magnetic fields between 0~T and $\pm 2.5$~T. The extracted magnetic-field variation of $\theta^{(3)}$ entered in Fig.~\ref{fig3}(c) differs from that of $M_{z}$. Unlike the LFR, which follows the relation $\theta^{(1)} \propto M_{z}$, the TFR is expressed by the relation $\theta^{(3)} \propto \arctan({\rm const}\cdot M_{z})$ according to Eq.~(\ref{eq:rot-n}).

In Figs.~\ref{fig2}(c) and \ref{fig3}(c) the value of $\arctan(|\chi_{yxxx}|/|\chi_{xxxx}|)$ is plotted and compared to the rotation angle $\theta^{(3)}$ directly measured in Figs.~\ref{fig2}(a) and \ref{fig3}(a). The agreement between the two data sets is obvious. This suggests that the approximation of ${\rm Re}({\rm i}\chi_{yxxx}/\chi_{xxxx})$ in Eq.~(\ref{eq:TFR}) by $|\chi_{yxxx}|/|\chi_{xxxx}|$, which neglects dichroic effects, is applicable for determining $\theta^{(3)}$. Because of the excellent agreement between the two data sets, we henceforth use the convenient approximation of $\theta^{(3)}$ via the third-order susceptibilities, instead of measuring it by an involved polarization analysis as in Figs.~\ref{fig2}(a) and \ref{fig3}(a).

\subsection{Comparing the microscopy of LFR and TFR}  \label{micro}

\begin{figure}
\includegraphics[width=\columnwidth,keepaspectratio,clip]{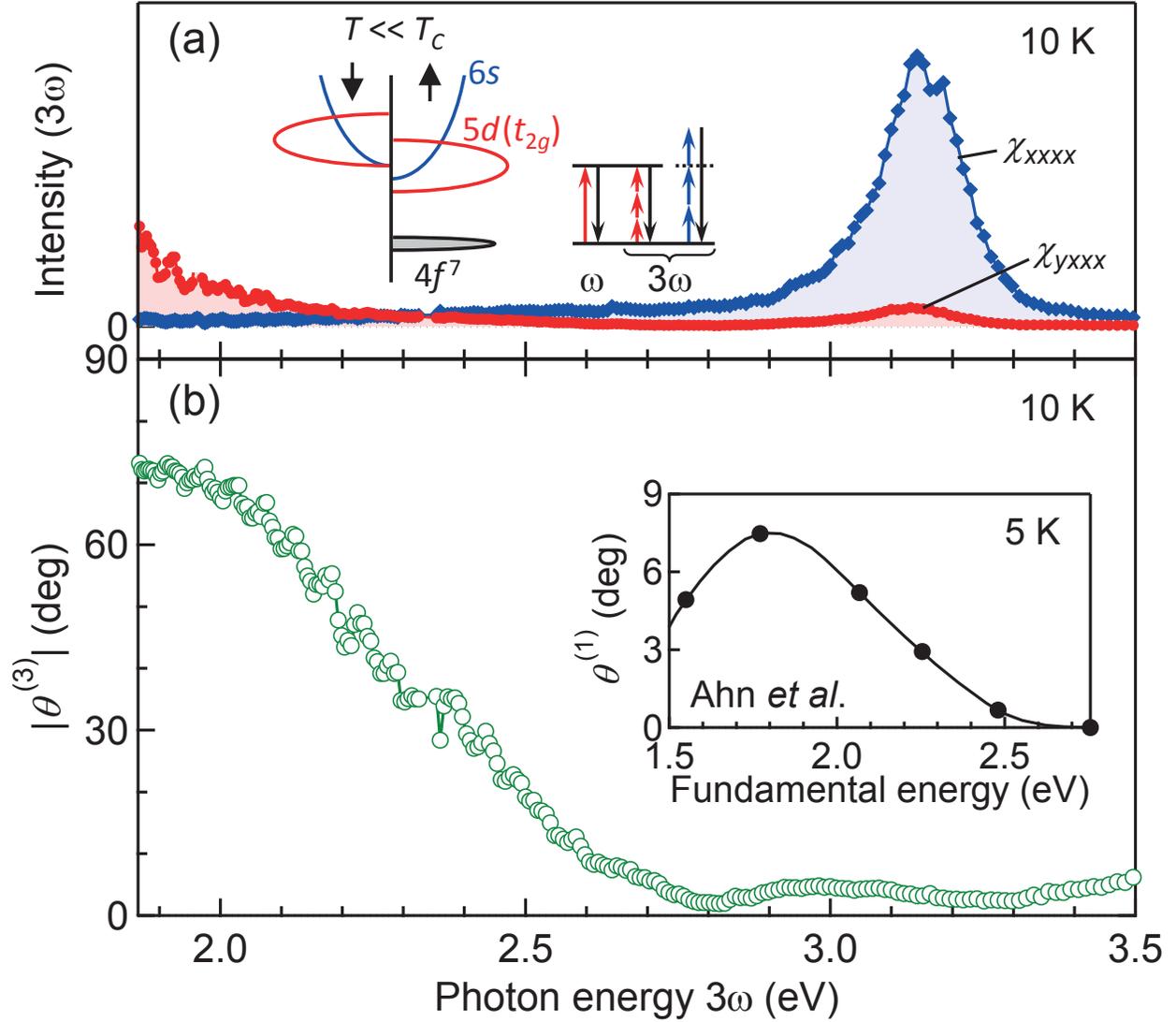}
\caption{(Color online) Spectral dependence of nonlinear susceptibilities and TFR of a EuO(001) film. (a) Spectrum of $\chi_{yxxx}$ and $\chi_{xxxx}$ at 10~K in a magnetic field $\mu_{0}H_{z}=+3$~T. The inset shows schematics of the spin-dependent electronic band structure of ferromagnetic EuO and of the optical transitions of the LFR and the TFR in the depicted spectral range. (b) Spectral dependance of the TFR derived from the data in (a) by $|\theta^{(3)}| \approx \arctan(|\chi_{yxxx}|/|\chi_{xxxx}|)$ (see text). Inset: Spectral dependence of the LFR at 5~K for a EuO film of 153~nm, taken from Ref.\protect~\onlinecite{Ahn_1967_IEEETM}.} \label{fig5}
\end{figure}

Finally, in order to disclose the microscopic mechanism of the giant TFR, its spectral origin has to be clarified. Therefore, Fig.~\ref{fig5}(a) shows the spectral dependence of the frequency-tripled signal for the magnetization-induced ($\chi_{yxxx}$) and the magnetization-insensitive ($\chi_{xxxx}$) susceptibilities and the estimated rotation $\theta^{(3)}$ at 10~K in a magnetic field $\mu_{0}H_{z}= +3$~T. While the slope of a resonance centered at $<1.9$~eV is present in $\chi_{yxxx}$ but not in $\chi_{xxxx}$, a pronounced peak around 3.1~eV is observed in both components. This corresponds to a specific resonance of $\theta^{(3)}$ at $<1.9$~eV, just like in the case of the LFR, as shown in Fig.~\ref{fig5}(b). The LFR is attributed to the transitions from the $4f^{7}$ ground state to the $4f^{6}5d^{1}(t_{2g})$ state of the Eu$^{2+}$ ion.\cite{Freiser_1968_HPA, Ahn_1967_IEEETM, Feinleib_1969_PRL, Wang_1986_HPA} It is caused by the spin polarization and the spin-orbit splitting of the $f$ and $d$ states involved in the optical transition. We associate the TFR to the same transition, yet as a {\it three-photon-resonant} excitation. This is reasonable because the selection rules for a one-photon transition are included in that of a three-photon transition. In contrast, the peak near 3.1~eV seems to involve a {\it two-photon-resonant} transition to the $4f^{6}5d^{1}(t_{2g})$ state followed by the transition via the third photon to the higher lying $5d/6s$ mixing state,\cite{Lafrentz_2010_PRB} as shown in the inset of Fig.~\ref{fig5}(a). This excitation does not contribute notably to the TFR. It reflects that the selection rules for the two-photon transition to the $4f^{6}5d^{1}(t_{2g})$ state are fundamentally different from the selection rules of the LFR and also that the $6s$ state with less magneto-optical activity is involved in the excitation process.

\section{Summary and Conclusion}

In summary, a giant third-order magneto-optical rotation termed TFR was observed in epitaxial ferromagnetic EuO films with a magnetic-field-induced out-of-plane magnetization. It results from the large spin polarization and the spin-orbit splitting of the states involved in the optical transition and reveals an inherent similarity to the LFR. However, the TFR is boosted by the ratio of the magnetic to the nonmagnetic tensor elements in the dielectric tensor $\hat{\epsilon}$. This ratio is much larger for the nonlinear than for the linear contributions. The giant TFR is particularly suitable for probing the magnetization of ultra-thin films and multilayers, because in contrast to the LFR, it is not affected by the reduction of the thickness of a material. In addition, the third-order Faraday rotation and the second-order Kerr rotation (commonly referred to as ``nonlinear magneto-optical Kerr rotation'') complement each other as probes for magnetism because of their different sensitivity to the symmetry.

\begin{acknowledgments}
This work was supported by the Alexander von Humboldt Foundation, by the SFB 608, the TRR 80 of the Deutsche Forschungsgemeinschaft, and by the AFOSR (Grant No. FA9550-10-1-0123).
\end{acknowledgments}



\newpage

\end{document}